\documentclass[aps,prc,twocolumn,showpacs,superscriptaddress,floatfix,10pt]{revtex4-1}
\usepackage{graphicx}
\usepackage{dcolumn}
\usepackage{bm}
\usepackage{amsfonts}

\begin{document}

\title{Microscopic study of the $^{132,124}$Sn+$^{96}$Zr reactions: dynamic excitation energy, \\
       energy-dependent heavy-ion potential, and capture cross section}

\author{V.E. Oberacker}
\author{A.S. Umar}
\affiliation{Department of Physics and Astronomy, Vanderbilt University, Nashville, Tennessee 37235, USA}
\author{J.A. Maruhn}
\affiliation{Institut f\"ur Theoretische Physik, Goethe-Universit\"at, D-60438 Frankfurt am Main, Germany}
\author{P.-G. Reinhard}
\affiliation{Institut f\"ur Theoretische Physik, Universit\"at Erlangen, D-91054 Erlangen, Germany}

\date{\today}


\begin{abstract}
We study reactions between neutron-rich $^{132}$Sn nucleus and
$^{96}$Zr within a dynamic microscopic theory at energies in the vicinity of the
ion-ion potential barrier peak, and we compare the properties to those of the stable system
$^{124}$Sn+$^{96}$Zr. The calculations are carried out on a three-dimensional lattice
using the density-constrained Time-Dependent Hartree-Fock method.
In particular, we calculate the dynamic excitation energy $E^*(t)$ 
and the quadrupole moment of the dinuclear system, $Q_{20}(t)$, during the
initial stages of the heavy-ion collision.
Capture cross sections for the two reactions are analyzed in terms of
dynamic effects and a comparison with recently measured data is given.
\end{abstract}
\pacs{21.60.-n,21.60.Jz}
\maketitle


\section{Introduction}
Heavy-ion reactions at radioactive ion beam (RIB) facilities allow us to form
new exotic neutron-rich nuclei and to study their physical properties.
Examples include experiments with neutron-rich $^{132}$Sn beams on targets
of $^{64}$Ni~\cite{Li03,Li08} and of $^{96}$Zr~\cite{V08}. Another
experimental frontier is the synthesis of superheavy nuclei in cold fusion reactions
involving spherical closed-shell $^{208}$Pb targets~\cite{Ho02} and in hot
fusion reactions with deformed actinide nuclei~\cite{Og10}.
These experiments present numerous challenges for a theoretical description,
in particular for dynamic microscopic theories.

At relatively large impact parameters, heavy-ion reactions are dominated by
deep inelastic collisions in which the nuclei make only
brief contact. The reaction products have mass and charge similar to
projectile and target, but the energy may be strongly damped. 
At smaller impact parameters, an intermediate dinuclear system is formed.
Entrance-channel heavy-ion potentials have been calculated in various 
models, including the macroscopic-microscopic method with five shape
parameters~\cite{II05}, and the energy density functional method
with extended Thomas-Fermi approximation~\cite{DN02}. Furthermore,
dynamical models~\cite{Fa04,NG09,AA09} show that if the dinuclear system
is able to move inside the saddle point, capture occurs which may lead
to the formation of a compound nucleus with compact shape. During capture the
energy of relative motion of the ions gets converted into intrinsic excitation energy
$E^*(t)$. If a compound nucleus is formed it will subsequently decay either by particle
evaporation or by fission.  On the other hand, if the dinuclear system separates
before crossing the saddle point, the reaction process is called quasifission.
In this case, the shape of the
dinuclear system is very elongated (large quadrupole moment). Experimentally,
it is possible to separate fusion-fission from quasifission by measuring the
angular anisotropy of the fragments.
In the collision of very heavy ions such as $^{132}$Sn+$^{96}$Zr considered
in this paper, studies of fusion reactions are complicated by
the competition with quasifission and fusion-fission events which hinder
the formation of evaporation residues.

The time-dependent Hartree-Fock (TDHF) theory provides a useful foundation for a
fully microscopic many-body theory of heavy-ion collisions in the vicinity of the
Coulomb barrier~\cite{Ne82,DS85}.
Partly because of the recent breakthroughs in microprocessor technology, it
has become feasible to perform TDHF calculations on a three-dimensional (3D) Cartesian grid
with no symmetry restrictions and with much
more accurate numerical methods.
At the same time the quality of effective interactions has also been substantially improved~\cite{CB98,BH03}.
These developments allow for the testing of the time-dependent mean-field approach to nuclear
reactions without any numerical uncertainties~\cite{UO06,GM08,DD-TDHF,KS10}.
The TDHF code used in these calculations utilizes the full Skyrme interaction, including all of the time-odd terms in the mean field Hamiltonian~\cite{EB75,UO06}.

During the past several years, we have developed the density-constrained (DC) TDHF
method (DC-TDHF) for calculating heavy-ion potentials~\cite{UO06a}. We have applied this method
to calculate fusion cross sections above and below the barrier for a
number of systems: The first application was for the $^{132}$Sn+$^{64}$Ni~\cite{UO06d,UO07a} system.
The fusion cross section at the lowest projectile energy has been re-measured~\cite{Li08}
and now agrees remarkably well with our calculations. 
We have also performed calculations for $^{64}$Ni+$^{64}$Ni~\cite{UO08a}
and for $^{16}$O+$^{208}$Pb\cite{UO09b}. In all these cases, we have found
very good agreement between the measured fusion cross sections and the DC-TDHF results.
Very recently, we have carried out a microscopic dynamical study of the astrophysical 
triple-$\alpha$ reaction to form a resonant state of $^{12}$C~\cite{UM10a} and a
study similar to the one presented here for reactions involving superheavy
formations~\cite{UO10a}, using the same approach.
 
In the present paper, we study reactions between the neutron-rich $^{132}$Sn nucleus and
$^{96}$Zr at energies in the vicinity of the ion-ion potential barrier peak, and we compare
observables to those of the stable system $^{124}$Sn+$^{96}$Zr. 
The dynamic microscopic calculations are carried out on a 3D Cartesian lattice
using both unrestricted TDHF and DC-TDHF methods. This is by far the heaviest neutron-rich
system we have investigated so far, and the microscopic numerical calculations with the
added DC are computationally very intensive.

This paper is organized as follows: in Section~II we summarize the Formalism
(DC-TDHF, dynamic excitation energy, capture cross section).
In Section~III numerical results are presented. In particular, we show contour
plots of the mass density of the dinuclear system and discuss the dynamic quadrupole moment 
$Q_{20}(t)$ during the initial stages of the collision.
We also calculate the heavy-ion interaction potential $V(R)$ and
demonstrate that in these very heavy systems the barrier height and
width increase dramatically with increasing beam energy.
Interaction barrier heights and positions are also deduced from unrestricted TDHF
runs. We examine the dynamic excitation energy $E^{*}(t)$ during the 
initial stages of the collision and compare it to the excitation energy
of the compound nucleus in its ground state, $E^{*}=E_{c.m.}+Q_{gg}$. Finally,
capture cross sections for the two reactions are analyzed in terms of
dynamic effects, and a comparison with recently measured capture-fission
data~\cite{V08} is given. The conclusions are presented in Section~IV.


\section{Formalism: DC-TDHF method, dynamic excitation energy, capture}

Recently, we have developed a method to extract ion-ion interaction potentials directly from
the TDHF time-evolution of the nuclear system.
In our DC-TDHF approach~\cite{UO06a}, the TDHF time-evolution takes place with no restrictions.
At certain times during the evolution the instantaneous density is used to
perform a static Hartree-Fock minimization while holding the total proton
and neutron density of the dinuclear system constrained
to be the instantaneous TDHF density. This provides us with the
TDHF dynamical path in relation to the multi-dimensional static energy surface
of the combined nuclear system.
In the DC-TDHF method the ion-ion interaction potential is given by
\begin{equation}
V(R)=E_{\mathrm{DC}}(R)-E_{\mathrm{A_{1}}}-E_{\mathrm{A_{2}}}\;,
\label{eq:vr}
\end{equation}
where $E_{\mathrm{DC}}$ is the DC energy at the instantaneous
separation $R(t)$, while $E_{\mathrm{A_{1}}}$ and $E_{\mathrm{A_{2}}}$ are the binding energies of
the two nuclei obtained with the same effective interaction.
The interaction potentials calculated with the DC-TDHF method incorporate
all of the dynamical entrance channel effects such as neck formation,
particle exchange, internal excitations, and deformation effects.
While the outer part of the potential barrier is largely determined by
the entrance channel properties, the inner part of the potential barrier
is strongly sensitive to dynamical
effects such as particle transfer and neck formation.

For the calculation of the ion-ion separation distance $R$ we use a hybrid method
as described in Ref.~\cite{UO09b}. At large distances
where a still visible neck allows us to identify two fragments we
compute it as distance of the center of mass of the ions. For more
compact configurations, we compute $R$ from the mass quadrupole moment $Q_{20}$
as $R=r_0\sqrt{|Q_{20}|}$ where $r_0$ is a scale factor to connect the definition smoothly 
to the large-distance region.

In heavy-ion reactions, the total capture cross section consists of the following
terms~\cite{V08}:
\begin{equation}
\sigma_{\mathrm{capt}}= \sigma_{\mathrm{ER}} + \sigma_{\mathrm{QF}}
+ \sigma_{\mathrm{FF}}\;,
\label{eq:sigma_capt_2}
\end{equation}
where $\sigma_{\mathrm{ER}}, \sigma_{\mathrm{QF}}, \sigma_{\mathrm{FF}}$ denote the evaporation
residue cross section, the quasifission cross section, and the fusion-fission cross section.
In the reaction of light and medium-heavy ions, the fission barriers of the pre-compound system are
so high that fission contributions are negligible. In this case we have
$\sigma_{\mathrm{capt}} \approx \sigma_{\mathrm{ER}}$. On the other hand, for the 
reaction of massive nuclei like $^{132}$Sn+$^{96}$Zr,
the pre-compound system is an excited state of the actinide nucleus $^{228}$Th with
a fission barrier of only about 6 MeV; thus we expect sizable fission contributions,
and the evaporation residue cross section is expected be be rather small. The number
of quasifission events increases dramatically with the product of the charge numbers
$Z_1 Z_2$ and with the orbital angular momentum $\ell$ in the entrance channel.
Another reason for the decreasing yield of ER formation is that
a heated and rotating CN may fission (fusion-fission).

Ion-ion interaction potentials calculated using DC-TDHF correspond to the
configuration attained during a particular TDHF collision. As mentioned above, for light and
medium mass systems as well as heavier systems for which fusion is
the dominant reaction channel, DC-TDHF
gives the fusion barrier with an appreciable but relatively small energy dependence.
On the other hand, for reactions involving massive systems fusion is
not the dominant channel at barrier top energies. Instead the system sticks
in some dinuclear configuration with possible break-up after exchanging a
few nucleons. The long-time evolution to break-up is beyond the scope of
TDHF due to the absence of quantum decay processes and transitions.
As we increase the energy above the barrier this phenomenon gradually
changes to the formation of a truly composite object. This is somewhat
similar to the \textit{extrapush} phenomenon discussed in phenomenological
models.
For this reason the energy dependence of the DC-TDHF ion-ion potential barriers
for these systems is not just due to the dynamical effects for the same final
configuration but actually represents different final configurations.

Theoretically, the calculation of the total capture cross section is
similar to the calculation of the fusion cross section
\begin{equation}
\sigma_{\mathrm{capt}} = \frac{\pi}{k^2} \sum_{L=0}^{\infty} (2L+1) T_L \ ,
\label{eq:sigma_capt_1}
\end{equation}
with the understanding that the ion-ion interaction potential used
in the calculations distinguishes the two events.
In practice, the potential barrier penetrabilities $T_L$ at $E_{\mathrm{c.m.}}$ energies
below and above the barrier are obtained by numerical integration of the
Schr\"odinger equation for the relative coordinate $R$ using
the well-established {\it Incoming Wave Boundary Condition} (IWBC) method~\cite{Raw64,HW07}.
This Schr\"odinger equation contains the heavy-ion potential
$V(R)$ given in Eq.~(\ref{eq:vr}) and the centrifugal potential.
In the IWBC calculations the summation over $L$ in Eq.~\ref{eq:sigma_capt_1} is
continued until the contribution becomes negligible to the total cross section.
As we shall discuss below, we can also determine the maximum value of $L$
by performing TDHF calculations for non-zero impact parameters.
The optimal way to study the problem would be to perform DC-TDHF calculations
for different $L$ values, however for heavy systems the computational cost
for doing this is very large. 
Finally, for the calculation of capture cross sections it is possible to use a coordinate dependent
effective mass $\mu(R)$ as described in Ref.~\cite{UO09b}. The effect of using a coordinate
dependent mass is to modify the inner part of the ion-ion potential, particularly at low subbarrier
energies. For the energies studied here we have found this effect to be very small for capture
cross sections.
\begin{figure}[!htb]
\includegraphics*[width=8.2cm]{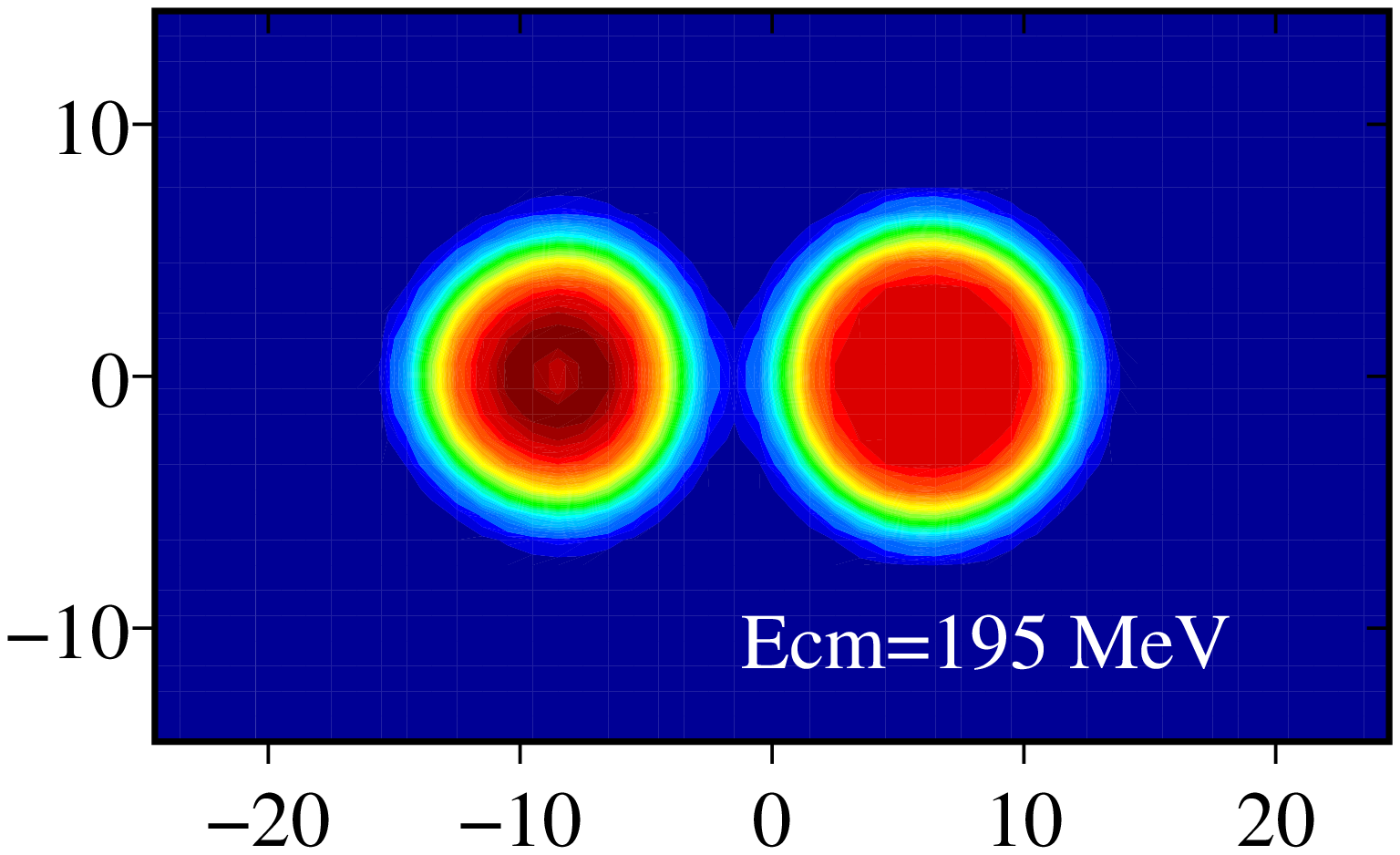}\vspace{-0.27in}
\includegraphics*[width=8.2cm]{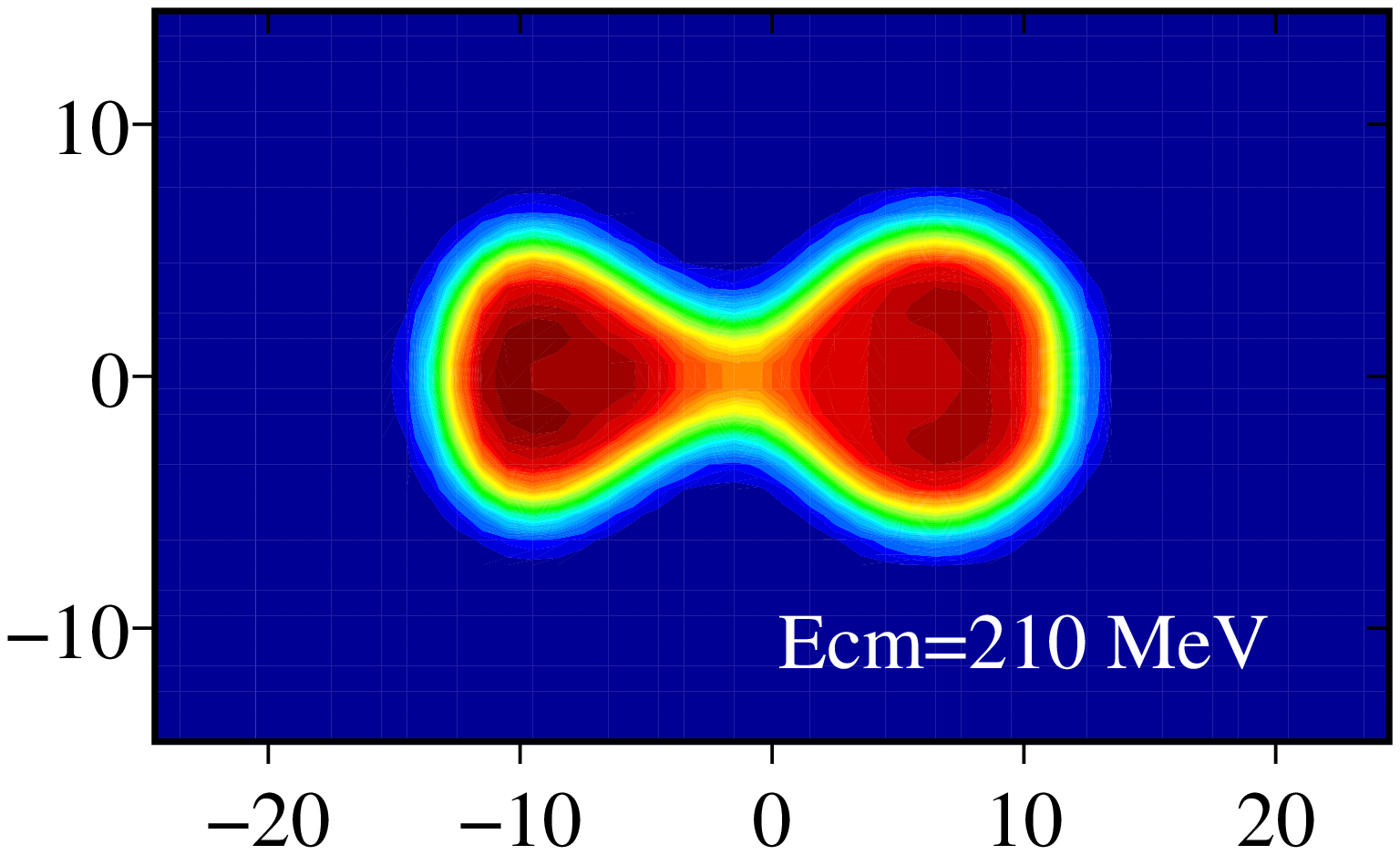}\vspace{-0.27in}
\includegraphics*[width=8.2cm]{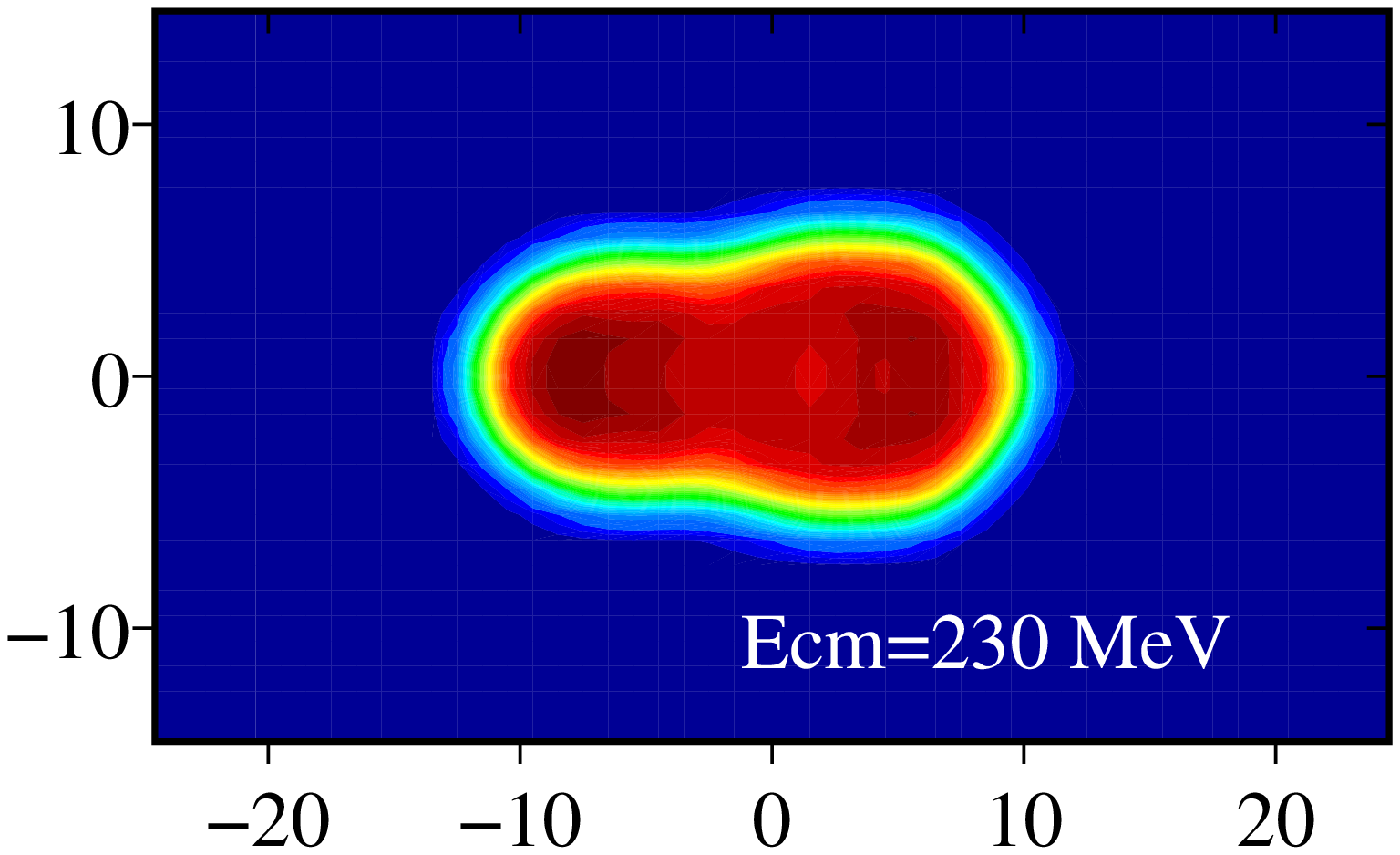}
\caption{\label{fig:dens} (color online) TDHF calculations for $^{132}$Sn+$^{96}$Zr. Contour
plots of the mass density at the distance of closest approach in a central collision, 
calculated at three different $E_\mathrm{c.m.}$ energies.}
\end{figure}

Taking up the strategy proposed in \cite{Cus85a},
we have also developed a new microscopic approach for calculating
dynamic excitation energies $E^{*}(t)$ of systems formed during heavy-ion
collisions~\cite{UOMR09}. For this purpose, we divide the conserved TDHF energy into
a collective and intrinsic part, and we 
assume that the collective part is primarily determined by
the density $\rho(\mathbf{r},t)$ and the current $\mathbf{j}(\mathbf{r},t)$.
Consequently, the excitation energy can be written in the form
\begin{equation}
E^{*}(t)=E_{\mathrm{TDHF}}-E_{\mathrm{coll}}\left(\rho(t),\mathbf{j}(t)\right)\;,
\label{eq:estar}
\end{equation}
where $E_\mathrm{TDHF}$ is the total energy of the dynamical system, which is a conserved quantity,
and $E_\mathrm{coll}$ represents the collective energy of the system. The collective energy
consists of two parts
\begin{equation}
E_{\mathrm{coll}}\left(t\right)= E_{\mathrm{kin}}\left(\rho(t),\mathbf{j}(t)\right)
 + E_{\mathrm{DC}}\left(\rho(t)\right)\;,
\end{equation}
where $E_{\mathrm{kin}}$ represents the kinetic part and is given by
\begin{equation}
E_{\mathrm{kin}}\left(\rho(t),\mathbf{j}(t)\right)=\frac{m}{2}\int\;{\rm d}^{3}r\;\mathbf{j}^2(t)/\rho(t)\;,
\label{eq:ekin}
\end{equation}
which is asymptotically equivalent to the kinetic energy of the
relative motion, $\frac{1}{2}\mu\dot{R}^2$, where $\mu$ is the
reduced mass and $R(t)$ is the ion-ion separation distance.
The energy $E_\mathrm{DC}$ is the lowest-energy state of all possible
TDHF states with the same density and is required to have zero excitation
energy. The dynamics of the ion-ion separation
$R(t)$ can be extracted from an unrestricted TDHF run. Using $E^{*}(t)$ and $R(t)$, we can deduce
the excitation energy as a function of the distance parameter, $E^{*}(R)$.


\section{Results}

The numerical calculations are carried
out on a 3D Cartesian lattice using the Basis-Spline collocation method to 
represent derivative operators with high accuracy.
For the $^{132,124}$Sn+$^{96}$Zr reactions studied here, the lattice spans
$50$~fm along the collision axis and $30-42$~fm in
the other two directions, depending on the impact parameter. The lattice spacing is 
$1.0$~fm in all directions. We utilize the full Skyrme interaction (SLy4)~\cite{CB98}
including all of the time-odd terms in the mean field Hamiltonian~\cite{UO06}, without
the c.m. correction as described in Ref.~\cite{UO09a}.
The two nuclei are placed at an initial separation of $22$~fm.
First we generate highly accurate static HF wave functions for the two nuclei on the
lattice, which are then boosted and time-propagated with a
time step $\Delta t = 0.4$ fm/c.
The computation of the dynamic excitation energy and the heavy-ion potential is
numerically very intensive, primarily due to the DC calculation.
In a typical DC-TDHF run, we utilize a few thousand time steps, and the DC is
applied every $20$ time steps.
To distinguish between deep-inelastic and capture reactions, we have also performed
several unrestricted TDHF runs for the $^{132}$Sn+$^{96}$Zr system above the barrier.
The numerical accuracy of the static binding energies and the deviation
from the point Coulomb energy in the initial state of the collision dynamics is on the order
of $50-200$~keV. The accuracy of the DC calculations is
commensurate with the accuracy of the static calculations.


\subsection{Dynamic quadrupole moment, Interaction Barrier, and Capture Barrier}

\begin{figure}[!htb]
\includegraphics*[width=8.6cm]{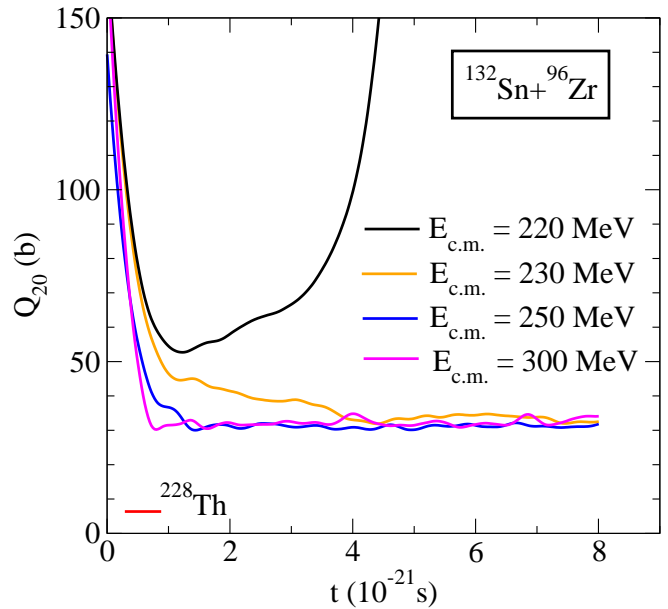}
\caption{\label{fig:q20t} (color online) Intrinsic mass quadrupole moment as a function of time.}
\end{figure}

In Fig.~\ref{fig:dens} we show contour plots of the mass density at the distance of closest
approach in a central collision.
These density plots reveal that at $E_{\mathrm{c.m.}}=195$ MeV the nuclear
surfaces barely touch; this energy corresponds to the interaction barrier.
At $E_{\mathrm{c.m.}}=210$ MeV we still notice a density configuration with
two separate cores. Only at energies $E_{\mathrm{c.m.}}=230$ MeV and above,
a single-core composite system emerges, albeit with very large elongation.
The large elongation of the composite system is readily apparent if
one plots the intrinsic mass quadrupole moment
\begin{equation}
Q_{20}(t)= \sqrt{\frac{5}{16 \pi}} \int d^3r \rho(\mathbf{r},t) (2 z^2 - x^2 - y^2)
\label{eq:Q20}
\end{equation}
as a function of time, see Fig.~\ref{fig:q20t}.
Also shown, for comparison, is the static quadrupole moment of the compound nucleus
$^{228}$Th which is more than three times smaller. Furthermore, the plot shows
that central collisions at energies $E_{\mathrm{c.m.}} \geq 230$ MeV lead to
capture while the nuclei bounce back from each other at $E_{\mathrm{c.m.}} = 220$ MeV
and below (deep-inelastic collision).

As we have discussed earlier, for systems leading to superheavy formation the evaporation residue cross section
is customarily represented in terms of the various phases of the reaction process as
\begin{equation}
\sigma_{\rm ER}=\sigma_{\rm capture}\cdot P_{\rm CN}\cdot P_{\rm survival}\;,
\end{equation}
where $\sigma_{\rm ER}$ denotes the evaporation-residue cross section for the superheavy system,
$\sigma_{\rm capture}$ is the capture cross section for the two-ions, $P_{\rm CN}$ is the probability of forming
a compound nucleus, and $P_{\rm survival}$ is the probability that this compound system survives
various breakup and fission events. The calculations presented here can only address the
capture cross section for these systems since the subsequent reaction possibilities are beyond
the scope of the TDHF theory. For most light systems, for which fusion is the dominant reaction
result, $\sigma_{\rm capture}$ and $\sigma_{\rm ER}$ are essentially the same and equal to the fusion
cross section, $\sigma_{\rm fusion}$. For reactions involving superheavy formations we instead have
\begin{equation}
\sigma_{\rm \rm capture}=\sigma_{\rm QF}+\sigma_{\rm FF}+\sigma_{\rm ER}\;,
\end{equation}
where $\sigma_{\rm QF}$ and $\sigma_{\rm FF}$ denote the quasi-fission and fusion-fission cross sections, respectively.
For these reactions the evaporation residue cross section, $\sigma_{\rm ER}$, is very small and therefore the
capture cross section is to a large extent equal to the sum of the two fission cross sections.
Furthermore, the distinction between deep-inelastic reactions  and quasi-fission is somewhat difficult and usually
achieved by setting windows for fragment masses of $A_{f} = A_{\rm CN}/2 \pm 20$ and on their kinetic energy.

\begin{figure}[!htb]
\includegraphics*[width=8.6cm]{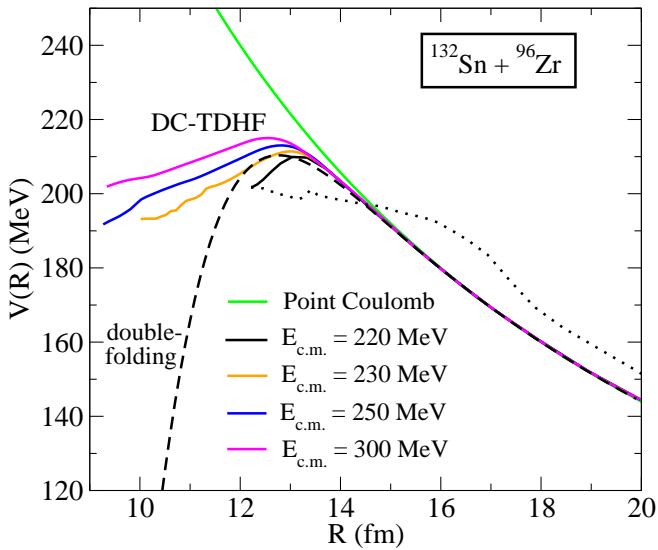}
\caption{\label{fig:pot} (color online) DC-TDHF calculations for the neutron-rich
system $^{132}$Sn+$^{96}$Zr. The potential barriers $V(R)$ at four $E_\mathrm{c.m.}$
energies are obtained using Eq.~\protect(\ref{eq:vr}). Also shown is the point
Coulomb potential. The phenomenological double-folding potential (dashed black curve)
is given for comparison.}
\end{figure}

In Fig.~\ref{fig:pot} we show the ion-ion potential $V(R)$
for a central collision of $^{132}$Sn+$^{96}$Zr, calculated at four different
$E_\mathrm{c.m.}$ energies using the DC-TDHF method. The dotted part of the
potential line calculated at $E_\mathrm{c.m.}=220$ MeV corresponds to the 
outgoing trajectory (the nuclei bounce back in a deep-inelastic collision).
Our results
demonstrate that in these very heavy systems the barrier height and
width increase dramatically with increasing energy $E_\mathrm{c.m.}$.
In fact, at higher energies the potential becomes almost flat. This is the
first n-rich system in which we have observed this behavior.
By contrast, DC-TDHF calculations for light ion systems such as $^{16}$O+$^{16}$O
show almost no energy-dependence even if we increase $E_\mathrm{c.m.}$ by a factor
of four~\cite{UOMR09}. Even in reactions between a light and a very heavy nucleus
such as $^{16}$O+$^{208}$Pb, we see only a relatively small energy dependence
of the barrier height and width~\cite{UO09b}.
For comparison, we have also plotted the phenomenological double-folding
potential~\cite{SL79,RO83b} which uses the ground state densities of the two nuclei
and keeps them frozen. This potential is energy-independent and
has been calculated using the M3Y effective NN interaction~\cite{BB77}
and static 2-D HFB densities. We observe that the double-folding potential
yields a potential barrier which is fairly similar to that of the DC-TDHF potential
at $E_\mathrm{c.m.}=230$ MeV; however, because of the frozen density approximation,
the potential exhibits too much attraction at smaller distances. Another difference
is that the ground state densities in the double folding method
correspond to well bound nucleons. The TDHF densities, on the other hand, have much
wider tails because they cover dynamically excited nucleons. This puts some part of the
excitation energy into the DC-TDHF energy and hence into the ion-ion potential.

\begin{figure}[!htb]
\includegraphics*[width=8.6cm]{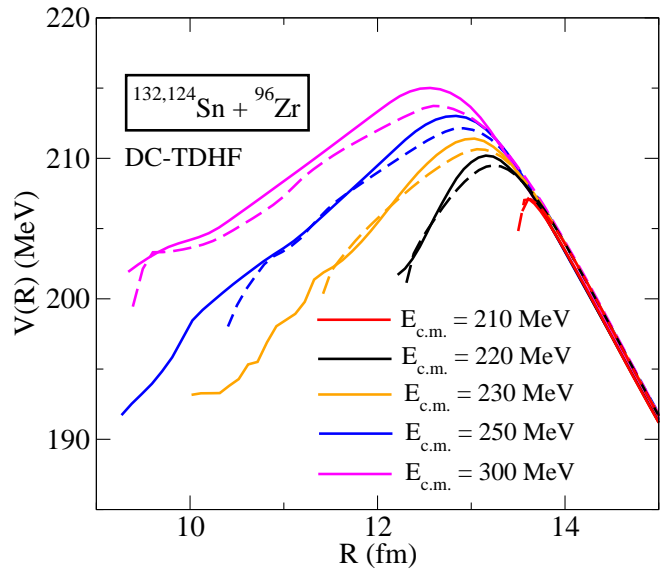}
\caption{\label{fig:pot1} (color online) Comparison of the heavy-ion barriers
for the neutron-rich system $^{132}$Sn+$^{96}$Zr (solid lines) and the stable system
$^{124}$Sn+$^{96}$Zr (dashed lines). The potential barriers are obtained with the DC-TDHF
method at five $E_\mathrm{c.m.}$ energies.}
\end{figure}

In Fig.~\ref{fig:pot1} we show the ion-ion potential barriers in detail. A comparison
is made between the neutron-rich system $^{132}$Sn+$^{96}$Zr and the stable system 
$^{124}$Sn+$^{96}$Zr. We find that the potential barriers of the neutron-rich system 
are systematically $1-2$ MeV higher than those of the stable system. 
We emphasize again that only the potential barriers calculated
at energies $E_{\mathrm{c.m.}} \geq 230$ MeV lead to a true composite system with
overlapping cores (capture) while the potential barriers calculated
at energies $E_{\mathrm{c.m.}} < 230$ MeV correspond to a dinuclear system where
both nuclei maintain separate cores (deep-inelastic collisions).

The interaction barrier $V_I$ is defined as the energy to bring the two colliding nuclei
into contact~\cite{V08}. This energy can be inferred from contour plots
of the TDHF mass density at the distance of closest approach in a central collision,
see Fig.~\ref{fig:dens}. For $^{132}$Sn+$^{96}$Zr, we find an interaction barrier height
$V_I=E_{\mathrm{c.m.}}=195$ MeV, and the corresponding distance of closest
approach $R_I=14.77$ fm.
In Table~\ref{table:barriers} we summarize the interaction barriers and ion-ion
potential barriers (capture barriers)
for the two systems and their corresponding positions in $R$-space. While the
DC-TDHF barriers are fairly similar, we observe large differences ($9$ MeV) in 
the interaction barriers of the two systems: the additional neutrons in $^{132}$Sn
give rise to a larger attractive potential which causes the nuclei to snap together
at lower energy.
\begin{table}[hbt!] 
\caption{\label{table:barriers}
Interaction barrier heights $V_I$ (energy to bring the two colliding nuclei
into contact) and barrier positions $R_I$
calculated with unrestricted TDHF at zero impact parameter. Also given are the 
energy-dependent ion-ion potential barrier heights $V_B$ and positions $R_B$ determined with
the DC-TDHF method.}
\begin{ruledtabular}
\begin{tabular}{ c c c c c}
Reaction & $V_I$ (MeV) & $R_I$ (fm) & $V_B$ (MeV) & $R_B$ (fm)  \\
\hline \\
$^{132}$Sn+$^{96}$Zr  & 195 & 14.77 & 211.4\footnotemark[1]  & 13.03\footnotemark[1] \\
                      &     &       & 215.0\footnotemark[2]  & 12.56\footnotemark[2] \\
\hline \\
$^{124}$Sn+$^{96}$Zr  & 204 & 14.05 & 210.6\footnotemark[1]  & 13.06\footnotemark[1] \\
                      &     &       & 213.7\footnotemark[2]  & 12.59\footnotemark[2]
\end{tabular}
\end{ruledtabular}
\footnotetext[1]{at $E_\mathrm{c.m.}=230$ MeV.}
\footnotetext[2]{at $E_\mathrm{c.m.}=300$ MeV.}
\end{table}
For TDHF collisions of light and medium mass systems as well as highly mass-asymmetric systems
fusion generally occurs immediately above the ion-ion potential barrier, while in heavier systems
there is an energy range above the barrier where capture does not occur.
The energy difference between the DC-TDHF potential barrier and the interaction barrier 
is the extrapush energy introduced by Swiatecki in a macroscopic model~\cite{Sw82};
in addition to the work presented here, this phenomenon has recently been studied for
heavy and nearly symmetric reaction partners using the TDHF method~\cite{SA09}. 


\subsection{Dynamic excitation energy $E^{*}(R)$}

In this subsection, we examine the dynamic excitation energy $E^{*}(R(t))$,
computed according to Eq.~(\ref{eq:estar}), during the 
initial stages of the collision. Of particular interest is the excitation 
energy at the capture point, $E^{*}_c$, which will influence the outcome of the reaction.
\begin{figure}[!htb]
\includegraphics*[width=8.6cm]{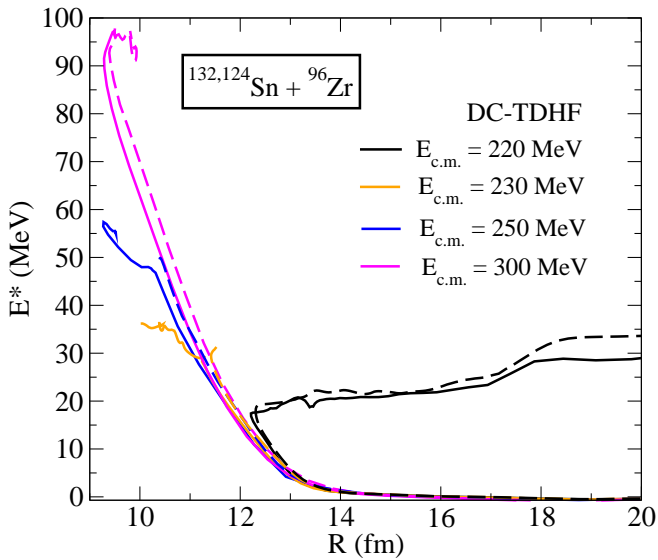}
\caption{\label{fig:estar_R} (color online) Pre-compound excitation energy as a function of the
internuclear distance $R$, calculated at zero impact parameter with the DC-TDHF method at four
$E_\mathrm{c.m.}$ energies. Compared are the neutron-rich system $^{132}$Sn+$^{96}$Zr (solid
lines) and the stable system $^{124}$Sn+$^{96}$Zr (dashed lines).}
\end{figure}
In Fig.~\ref{fig:estar_R} the pre-compound excitation energies are shown as a function of the internuclear distance $R$; this represents our first microscopic calculation for neutron-rich systems.
When the two nuclei are far apart, the excitation energy is zero (this provides a good
test for the numerical accuracy of the DC-TDHF calculation). As the two ions approach each
other the excitation energy increases rapidly and reaches values between 30 - 90 MeV
for the given range of c.m. energies. It is interesting to note that at
$E_\mathrm{c.m.}= 220$ MeV TDHF theory predicts that the two ions bounce back
despite the fact that they are almost 10 MeV above the corresponding potential barrier,
i.e. at this energy we have predominantly deep-inelastic reactions rather than capture.
This is due to the fact that a large part of the incoming c.m. energy was converted
to internal excitation $E^*$ such that the collective energy does not suffice any more
to surmount the barrier.
This feature was also shown in the figure for the corresponding heavy-ion interaction
potential (dotted line in Fig.~\ref{fig:pot}). 

\begin{figure}[!htb]
\includegraphics*[width=8.6cm]{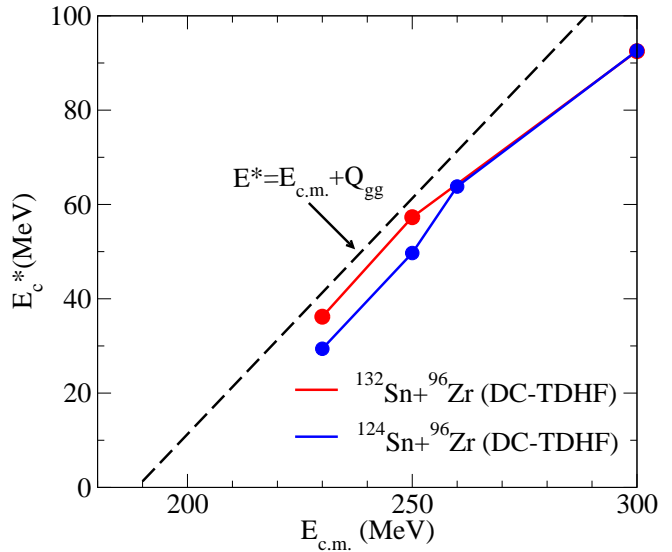}
\caption{\label{fig:estar_ecm} (color online) Pre-compound excitation energy 
at the capture point, $E^{*}_c$, as function of the c.m. energy,
as predicted by DC-TDHF for a central collision.}
\end{figure}

In Fig.~\ref{fig:estar_ecm} we show the excitation energy $E^{*}_c$ at the capture point
as a function of $E_\mathrm{c.m.}$. The capture point is defined as the distance $R$
inside the barrier region where the collective kinetic energy, Eq.~(\ref{eq:ekin}), becomes zero.
Because of the very elongated shape of the dinuclear composite system at the
capture point,  $E^{*}_c$ is systematically lower than one would expect for
the compound nuclei $^{228,220}$Th in their ground state (dashed line in Fig.~\ref{fig:estar_ecm})
\begin{equation}
E^{*}=E_{c.m.}+Q_{gg} \ .
\label{eq:Qgg}
\end{equation}
The last expression can be derived from reaction kinematics. The Q-values are obtained from measured
binding energies of the reaction partners and the compound nuclei. For the two systems
considered here the Q-values are almost identical: $Q_{gg}=-188.7$ MeV for $^{132}$Sn+$^{96}$Zr
and $Q_{gg}=-188.4$ MeV for $^{124}$Sn+$^{96}$Zr, hence we have drawn only one curve
for both systems (dashed line in Fig.~\ref{fig:estar_ecm}).
We observe that the excitation energy $E^{*}_c$ at the capture point is somewhat
lower for the $^{124}$Sn+$^{96}$Zr system in the energy range $E_\mathrm{c.m.}=230-250$ MeV;
at higher $E_\mathrm{c.m.}$ energies, their excitation energies are almost identical.
In this context, we would like to mention that recent microscopic calculations~\cite{PN09} 
have shown that the temperature (excitation energy) of the actinide compound nuclei
will strongly influence the height of their fission barriers.


\subsection{Capture and deep-inelastic cross section for $^{132}$Sn+$^{96}$Zr}

\begin{figure}[!htb]
\includegraphics*[width=8.6cm]{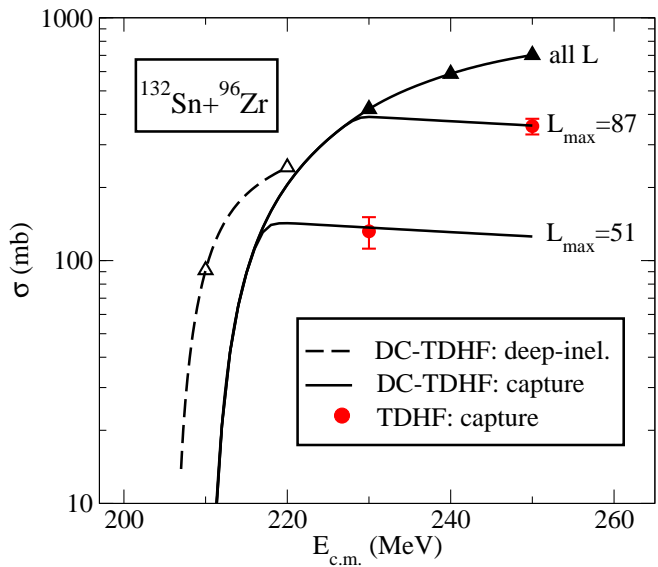}
\caption{\label{fig:capt132_v2} (color online) Total cross section for capture (solid line) and for
deep-inelastic reactions (dashed line) for the neutron-rich system
$^{132}$Sn+$^{96}$Zr calculated with the DC-TDHF method as function of
$E_{\mathrm{c.m.}}$. Total capture cross sections predicted
by unrestricted TDHF calculations are also given (red dots with error
bars). For details, see the text.}
\end{figure}

Previously, we have studied heavy-ion fusion of the neutron-rich system
$^{132}$Sn+$^{64}$Ni using the DC-TDHF method~\cite{UO06d,UO07a}.
In that case, the fission barrier of the compound system
is so high that the fission contribution is negligible at energies
near the Coulomb barrier.
By contrast, the compound nuclei for the systems studied in the present paper,
$^{132,124}$Sn+$^{96}$Zr, are the actinides $^{228,220}$Th
with a fission barrier of only about $6$ MeV. We therefore expect sizable fission
competition, and the evaporation residue cross section will be quite small.
Depending upon beam energy and impact parameter, the dominant reaction channels
are deep-inelastic and capture reactions. In general, central collisions and
collisions with relatively small impact parameter result in capture (one fragment
in the exit channel), while at larger impact parameters the system tends to
disintegrate into two fragments after some mass and charge transfer (deep-inelastic 
reactions). Regarding the capture channel, the composite system will 
eventually decay by nucleon and photon emission or by fission. This
long-time evolution of the composite system is beyond the scope of TDHF
due to the absence of quantum decay processes and transitions.
In Fig.~\ref{fig:capt132_v2} we show total capture and deep-inelastic
cross sections for the neutron-rich system $^{132}$Sn+$^{96}$Zr.
Let us first discuss the results obtained with the DC-TDHF method,
which can be used at energies $E_{\mathrm{c.m.}}$ below and above
the potential barriers. From a comparison of the heavy-ion potentials
in Fig.~\ref{fig:pot1} with the nuclear density distributions at the
distance of closest approach, see Fig.~\ref{fig:dens}, we conclude that
only potentials calculated
at energies $E_{\mathrm{c.m.}} \geq 230$ MeV lead to a true composite system with
overlapping cores, i.e. a capture reaction. By contrast, the heavy-ion potentials calculated
at energies $E_{\mathrm{c.m.}} < 230$ MeV correspond to a dinuclear system where
both nuclei maintain separate cores, i.e. deep-inelastic reactions.
The DC-TDHF capture cross section at $E_{\mathrm{c.m.}}=230$ MeV has been calculated from
the energy-dependent heavy-ion potential at $230$ MeV.
Similar calculations were carried out at energies $E_{\mathrm{c.m.}}=240$ and $250$ MeV.
These capture cross sections are marked by filled triangles, and the solid line
represents an interpolation between the data points. 
No restrictions were applied to the sum of partial waves $L$ in Eq.~(\ref{eq:sigma_capt_1}),
and this curve is therefore marked ``all $L$''. 
The capture cross sections
at energies below $230$ MeV were obtained from the heavy-ion potential at $230$ MeV
because this is the lowest potential barrier that leads to capture. 
The deep-inelastic cross section at $E_{\mathrm{c.m.}}= 220$ MeV was calculated
using the energy-dependent heavy-ion potential at $220$ MeV.
A similar calculation was carried out at energy $E_{\mathrm{c.m.}}=210$ MeV.
These deep-inelastic cross sections are marked by open triangles, and the dashed line
represents an interpolation between these data points. The capture cross sections
at energies below $210$ MeV were calculated from the heavy-ion potential at $210$ MeV
because this is the lowest potential barrier predicted by DC-TDHF. 
\begin{figure}[!htb]
\includegraphics*[width=8.6cm]{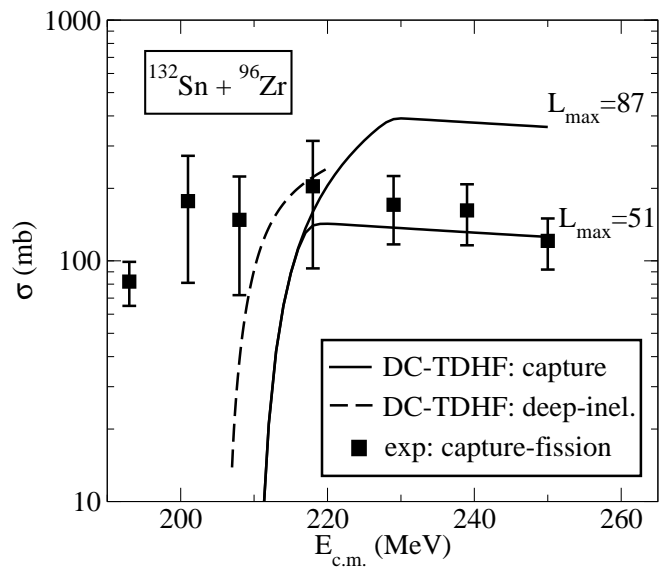}
\caption{\label{fig:capt_fiss132} Total cross section for capture (solid line) and for
deep-inelastic reactions (dashed line) for the neutron-rich system
$^{132}$Sn+$^{96}$Zr calculated with the DC-TDHF method as function of
$E_{\mathrm{c.m.}}$. The experimental
capture-fission cross sections are taken from Ref.~\cite{V08}.}
\end{figure}

At energies above the potential barriers (no barrier tunneling), we have also carried out unrestricted
TDHF runs with impact parameters in the range of $0-4$ fm. By examining the density
contours as a function of time, one can easily distinguish between capture
events (one fragment in the exit channel) and deep-inelastic reactions (two fragments).
At $E_{\mathrm{c.m.}}=250$ MeV we find that impact parameters $b=0-3.25$ fm result
in capture, while impact parameters $b \geq 3.50$ fm lead to deep-inelastic reactions.
Using the sharp cut-off model, the capture cross section is given by
$\sigma_{\mathrm{capt}} = \pi b_{max}^2$ with $b_{max}=3.375 \pm 0.125$ fm.
This cross section is shown by a red dot in Fig.~\ref{fig:capt132_v2}, with
the corresponding theoretical error bar arising from the impact parameter spacing.
The impact parameter $b_{max}=3.375$ fm corresponds to an orbital angular momentum
quantum number $L_{max}=87$. If we use this angular momentum cut-off in the IWBC
method, we obtain the curve labeled $L_{max}=87$; as we can see, both methods
yield the same capture cross section. We have carried out a similar calculation at
$E_{\mathrm{c.m.}}=230$ MeV which yields $b_{max}=2.05 \pm 0.15$ fm and $L_{max}=51$.
The corresponding TDHF and DC-TDHF capture cross sections are also plotted in
Fig.~\ref{fig:capt132_v2}.
\begin{figure}[!htb]
\includegraphics*[width=8.6cm]{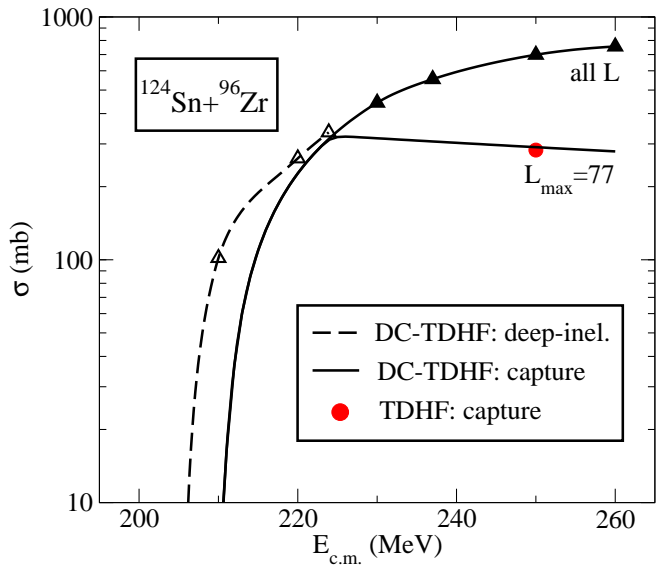}
\caption{\label{fig:capt124_v2}  (color online) Total cross section for capture (solid line) and for
deep-inelastic reactions (dashed line) for the stable system
$^{124}$Sn+$^{96}$Zr calculated with the DC-TDHF method as function of
$E_{\mathrm{c.m.}}$. The red dot represents the total capture cross section
at $E_{\mathrm{c.m.}}=250$ MeV predicted by an unrestricted TDHF run.
For details, see text.}
\end{figure}

In Fig.~\ref{fig:capt_fiss132} we compare the DC-TDHF cross sections for deep-inelastic and
capture reactions to experimental capture-fission cross sections measured at HRIBF
with a radioactive $^{132}$Sn beam~\cite{V08}. Because the fission probability $P_{\mathrm{fiss}} \leq 1$,
our \emph{calculated capture cross sections} should be regarded as an \emph{upper limit}
for the measured capture-fission cross sections.
According to our unrestricted TDHF calculations, the dominant reaction channels
at energies $E_{\mathrm{c.m.}} < 230$ MeV are the deep-inelastic and quasi-elastic
channels. In fact, our density plots in Fig.~\ref{fig:dens} reveal that at the lowest measured energy
$E_{\mathrm{c.m.}}=195$ MeV the nuclear surfaces barely touch. Any fission from
such an event would have to arise from sub-barrier neutron-transfer and
should be negligible compared to capture fission at higher energies.
We therefore make the conjecture that the bulk of the low-energy experimental data
in Fig.~\ref{fig:capt_fiss132} represent deep-inelastic and quasi-elastic
events masquerading as capture-fission. Indeed, 
because of the limited mass resolution in the HRIBF experiments~\cite{V08}
it has been difficult to separate the DIC component from capture-fission.
Further experiments with an improved fission fragment detector are planned~\cite{Sh09}.


\subsection{Capture and deep-inelastic cross section for $^{124}$Sn+$^{96}$Zr}

In Fig.~\ref{fig:capt124_v2} we examine the properties of the 
stable system $^{124}$Sn+$^{96}$Zr. 
Plotted are total cross sections for capture and for deep-inelastic reactions
calculated with the DC-TDHF method as function of the c.m. energy.
Like in the corresponding neutron-rich system, only the heavy-ion potentials calculated
at energies $E_{\mathrm{c.m.}} \geq 230$ MeV lead to capture while potentials
at lower energies are associated with deep-inelastic channels.
We have also carried out unrestricted TDHF runs for this system at
$E_{\mathrm{c.m.}}=250$ MeV. At impact parameter $b=3.5$ fm we find
a deep-inelastic reaction, and at $b=3.0$ fm we obtain capture.
Again, using the sharp cut-off model with $b_{max}=3.0$ fm we obtain
the cross section shown by a red dot in Fig.~\ref{fig:capt124_v2}.
This impact parameter corresponds to an orbital angular momentum
quantum number $L_{max}=77$. If we use this angular momentum cut-off in the DC-TDHF
method, we obtain the curve labeled $L_{max}=77$; as we can see, both methods
yield the same capture cross section.
\begin{figure}[!htb]
\includegraphics*[width=8.6cm]{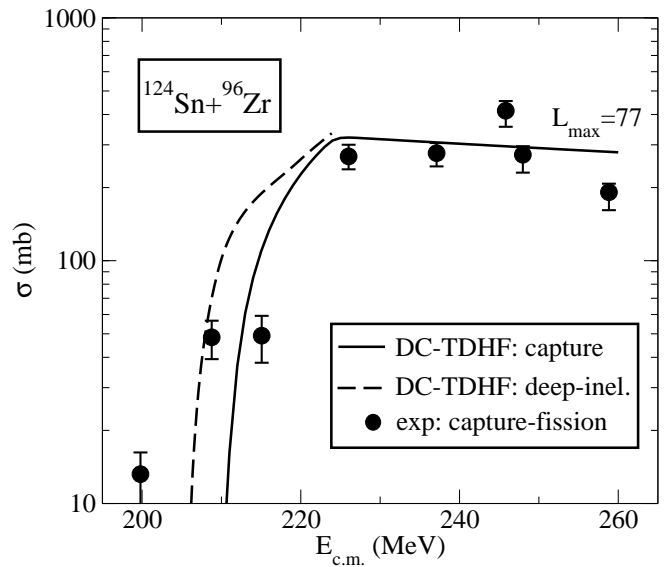}
\caption{\label{fig:capt124} Total cross section for capture (solid line) and for
deep-inelastic reactions (dashed line) for the stable system
$^{124}$Sn+$^{96}$Zr calculated with the DC-TDHF method as function of
$E_{\mathrm{c.m.}}$. The experimental capture-fission
cross sections are taken from Ref.~\cite{V08}.}
\end{figure}

In Fig.~\ref{fig:capt124} we compare the DC-TDHF cross sections for deep-inelastic and
capture reactions to experimental capture-fission cross
sections measured at HRIBF. Again, our calculated capture cross sections
represent an upper limit for the measured capture-fission data.
The agreement between theory and experiment is quite remarkable
in view of the fact that we employ a fully microscopic theory
based on a given energy functional, with
no adjustable parameters related to the capture process.


\section{Conclusions}

In this paper we have studied deep-inelastic and capture reactions for the neutron-rich 
system $^{132}$Sn+$^{96}$Zr at energies in the vicinity of the  barrier. This is by far
the heaviest neutron-rich system we have investigated using both unrestricted TDHF
and DC-TDHF methods. To elucidate any special properties neutron-rich systems might
possess, we have compared a number of observables to those of the stable
system $^{124}$Sn+$^{96}$Zr. The dynamic microscopic calculations are carried out
on a 3D Cartesian lattice, and they require a large amount of CPU time,
particularly with the added DC method.

A contour plot of the mass density of the dinuclear system shows clearly a 
transition from two separate cores at lower energies to a shape configuration
with overlapping cores or a single-core at energies $E_{\mathrm{c.m.}} \geq 230$ MeV.
A study of the dynamic quadrupole moment $Q_{20}(t)$ shows that even at
$E_{\mathrm{c.m.}}=300$ MeV, the intrinsic quadrupole moment is $3$ times larger
than that of the deformed compound nucleus $^{228}$Th during the initial stages
of the collision.
We also calculate the heavy-ion interaction potential $V(R)$, and we 
demonstrate that in these very heavy systems the barrier height and
width increase dramatically with increasing beam energy $E_\mathrm{c.m.}$.
We find that the potential barriers of the neutron-rich system $^{132}$Sn+$^{96}$Zr
are systematically $1-2$ MeV higher than those of the stable system. 
By contrast, we observe large differences ($9$ MeV) in the interaction barriers
of the two systems which can be deduced from unrestricted TDHF
runs. We then examine the dynamic excitation energy $E^{*}(t)$ during the 
initial stages of the collision and compare it to the expression
$E^{*}=E_{c.m.}+Q_{gg}$ (deduced from reaction kinematics) which 
assumes that the combined system is in its ground state. Finally,
capture cross sections for the two reactions are analyzed in terms of
dynamic effects, and a comparison with recently measured capture-fission
data is given.

One of the major open questions in the reactions of neutron-rich nuclei
is the dependence of the  barrier on isospin $T_z = (Z-N)/2$. 
To reveal possible systematic trends requires additional theoretical
and experimental studies with a wide variety of projectile and target
combinations which are expected to become available at current and future 
RIB facilities. To be able to pin down the isospin dependence in a 
fully microscopic theory, it is probably best to choose collision partners
which are as simple as possible: projectile and target nuclei should be
spherical in their ground state, and the compound nucleus should have a high fission barrier
so that the fission component can be ignored (at least at lower beam energies). A desirable 
reaction system of this kind appears to be $^{132}$Sn+$^{40,48,54}$Ca, and
we are planning to investigate these reactions in the future.


\begin{acknowledgments}
This work has been supported by the U.S. Department of Energy under Grant No.
DE-FG02-96ER40963 with Vanderbilt University, and by the German BMBF
under Contracts No. 06F131 and No. 06ER142D. 
\end{acknowledgments}



\begin{thebibliography}{99}
\bibitem{Li03}  J. F. Liang \textit{et al.}, Phys. Rev. Lett. {\bf 91}, 152701 (2003); {\bf 96}, 029903(E) (2006).
\bibitem{Li08}  J. F. Liang. D. Shapira, C. J. Gross, R. L. Varner, J. R. Beene, P. E. Mueller,
                and D. W. Stracener, Phys. Rev. C {\bf 78}, 047601 (2008).
\bibitem{V08}   A. M. Vinodkumar \textit{et al.}, Phys. Rev. C {\bf 78}, 054608 (2008).
\bibitem{Ho02}  S. Hofmann \textit{et al.}, Eur. Phys. J. A \textbf{14}, 147 (2002).
\bibitem{Og10}  Yu. Ts. Oganessian \textit {et al.}, Phys. Rev. Lett. \textbf {104}, 142502 (2010).
\bibitem{II05}  T. Ichikawa, A. Iwamoto, P. M\"oller, and A.J. Sierk, Phys. Rev. C {\bf 71}, 044608 (2005).
\bibitem{DN02}  V. Yu. Denisov and W. N\"orenberg, Eur. Phys. J. A {\bf 15}, 375 (2002).
\bibitem{Fa04}  G. Fazio, \textit{et. al.}, Eur. Phys. J. A \textbf{19}, 89 (2004).
\bibitem{NG09}  A. K. Nasirov, G. Giardina, G. Mandaglio, M. Manganaro, F. Hanappe, S. Heinz, 
                S. Hofmann, A. I. Muminov, W. Scheid, Phys. Rev. C \textbf{79}, 024606-10 (2009).
\bibitem{AA09}  G. G. Adamian, N. V. Antonenko , and W. Scheid,  Eur. Phys. J. A \textbf{41}, 235 (2009).
\bibitem{Ne82}  J. W. Negele, Rev. Mod. Phys. \textbf{54}, 913 (1982).
\bibitem{DS85}  K. T. R. Davies, K. R. Sandhya Devi, S. E. Koonin, and M. R. Strayer,
                page 3 in ``Treatise on Heavy--Ion Physics, Vol. 3 Compound System Phenomena''
                edited by D. A. Bromley, Plenum Press, New York, 1985.
\bibitem{CB98}  E. Chabanat, P. Bonche, P. Haensel, J. Meyer and R. Schaeffer, Nucl. Phys. \textbf{A635}, 231 (1998);
                \textbf{A643}, 441(E) (1998).
\bibitem{BH03}  M. Bender, P.-H. Heenen, and P.-G. Reinhard,
                Rev. Mod. Phys. \textbf{75}, 121 (2003).
\bibitem{UO06}  A. S. Umar and V. E. Oberacker, Phys. Rev. C \textbf{73}, 054607 (2006).
\bibitem{GM08}  Lu Guo, J. A. Maruhn, P.-G. Reinhard, and Y. Hashimoto, Phys. Rev. C \textbf{77},
                041301(R) (2008).
\bibitem{DD-TDHF} Kouhei Washiyama and Denis Lacroix, Phys. Rev. C {\bf 78}, 024610 (2008).
\bibitem{KS10} David J. Kedziora and C\'edric Simenel, Phys. Rev. C {\bf 81}, 044613 (2010).
\bibitem{EB75} Y. M. Engel, D. M. Brink, K. Goeke, S. J. Krieger, and D. Vautherin,
               Nucl. Phys. A \textbf{249}, 215 (1975).
\bibitem{UO06a} A. S. Umar and V. E. Oberacker, Phys. Rev. C \textbf{74}, 021601(R) (2006).
\bibitem{UO06d} A. S. Umar and V. E. Oberacker, Phys. Rev. C \textbf{74}, 061601(R) (2006).
\bibitem{UO07a} A. S. Umar and V. E. Oberacker, Phys. Rev. C {\bf 76}, 014614 (2007).
\bibitem{UO08a} A. S. Umar and V. E. Oberacker, Phys. Rev. C \textbf{77}, 064605 (2008).
\bibitem{UO09b} A. S. Umar and V. E. Oberacker, Eur. Phys. J. A \textbf{39}, 243 (2009).
\bibitem{UM10a} A. S. Umar, J. A. Maruhn, N. Itagaki, and V.E. Oberacker, Phys. Rev. Lett. \textbf {104}, 212503 (2010).
\bibitem{UO10a} A. S. Umar, V. E. Oberacker, J. A. Maruhn, and P.-G. Reinhard, Phys. Rev. C \textbf{81} 064607 (2010).
\bibitem{UO06b} A. S. Umar and V. E. Oberacker, Phys. Rev. C \textbf{74}, 061601(R) (2006).
\bibitem{Raw64} G. H. Rawitscher, Phys. Rev. 135, 605 (1964).
\bibitem{HW07}  K. Hagino and Y. Watanabe, Phys. Rev. C {\bf 76}, 021601(R) (2007).
\bibitem{Cus85a} R. Y. Cusson, P.-G. Reinhard, M. R. Strayer, J. A. Maruhn, and W. Greiner,
                 Z. Phys. A \textbf{320}, 475 (1985).
\bibitem{UOMR09} A. S. Umar, V. E. Oberacker, J. A. Maruhn, and P.-G. Reinhard,
        Phys. Rev. C  \textbf{80}, 041601(R) (2009).
\bibitem{UO09a} A. S. Umar and V. E. Oberacker, J. Phys. G: Nucl. Part. Phys. \textbf{36}, 025101 (2009).
\bibitem{SL79} G. R. Satchler and W. G. Love, Phys. Rep. \textbf{55}, 183 (1979).
\bibitem{RO83b} M. J. Rhoades-Brown, V. E. Oberacker, M. Seiwert, and W. Greiner,
                Z. Phys. A \textbf{310}, 287 (1983).
\bibitem{BB77} G. Bertsch, J. Borysowicz, H. McManus, and W. G. Love,
               Nucl. Phys. A \textbf{284}, 399 (1977).
\bibitem{Sw82}  W. Swiatecki, Nucl. Phys. A \textbf{376}, 275 (1982).
\bibitem{SA09}  C\'edric Simenel, Beno\^{\i}t Avez, C\'edric Golabek, (http://arxiv.org/abs/0904.2653v1).
\bibitem{PN09}  J. C. Pei, W. Nazarewicz, J. A. Sheikh, and A. K. Kerman, Phys. Rev. Lett. \textbf{102}, 192501 (2009).
\bibitem{Sh09} D. Shapira, ORNL, private communication.
\end{thebibliography}
\end{document}